\documentstyle[12pt,aasms4,epsfig]{article} 
 
\begin{document} 
 
\title{On the robustness of the microlensing optical depth as a
measure of MACHO density} 
 
\author{E. J. Kerins}

\affil{URA CNRS 1280, Observatoire de Strasbourg, 11 Rue de
l'Universit\'e, F-67000 Strasbourg, France}

\authoraddr{Email: kerins@wirtz.u-strasbg.fr}

\newcommand{\sm}{\mbox{M}_{\sun}} 
\newcommand{\tst}{\textstyle} 
\newcommand{\be}{\begin{equation}} 
\newcommand{\ee}{\end{equation}} 
 
\begin{abstract} 
The optical depth is widely used in Galactic microlensing studies as a
means to determine the density of MACHOs, since in theory it depends
only upon their spatial distribution and is therefore less
model-dependent than other microlensing observables.  However, since
the measured optical depth is restricted to that of events with
measurable timescales, inferences of total MACHO density are dependent
on the assumed timescale distribution. Using the 2-year LMC results of
the MACHO collaboration, we illustrate this point by showing how, for
an assumed isothermal halo, the inferred MACHO halo fraction
as determined from optical depth estimates depends upon MACHO mass.

The analysis highlights the following conclusions: (1) The MACHO
density inferred from optical depth measurements depends on the
assumed MACHO mass function for a given Galactic distribution
function; (2) without extra information on the MACHO mass function,
such as can be obtained from the rate-timescale distribution, optical
depth measurements can provide a lower limit but not an upper limit on
the MACHO density for a given Galactic distribution function; (3) a
comparison between the inferred total optical depth of different
Galactic models, or of different components in a multi-component
Galactic model, requires knowledge of the underlying timescale
distribution for each model or component.

For our assumed Galactic halo model we find excellent agreement
between our model-dependent lower-limit halo fraction of $f > 0.3$
($84\%$ confidence) or $f> 0.15$ ($97.5\%$ confidence), and the MACHO
collaboration's own model-independent optical depth lower-limit
estimates. MACHO's optical depth upper limits are consistent with the
minimum value of the upper limit derived for our assumed model.
\end{abstract} 
 

\section{Introduction} \label{s1} 

Several microlensing experiments are undertaking searches to detect
compact halo dark matter (MACHOs), as well as other low-luminosity
stellar populations, following the suggestion of Paczy\'nski
(\cite{pac86}). One of the principal quantities which characterises the
microlensing properties of a particular MACHO population is the
optical depth $\tau$. This quantity determines the average number of
microlensing events in progress at any instant in time per background
source star. In its simplest form it is given by
   \be
      \tau = \int_{0}^{L} \frac{\pi R_{\rm e}}{m} \rho(x) \, {\rm d}x,
      \label{e1}
   \ee
where $x$ is the distance along the line of sight between the observer and 
MACHO, $L$ is the observer--source distance, $m$ is the MACHO mass,
$\rho$ is the MACHO mass density at $x$ and 
   \be
      R_{\rm e} = \sqrt{\frac{4 G m x(L-x)}{c^2 L}} \label{e2}
   \ee
is the Einstein radius. Equation~(\ref{e1}) is valid if all sources are at
distance $L$, otherwise one needs to further integrate equation~(\ref{e1})
over the spatial distribution of sources (\cite{kir94}).

So far, theoretical predictions have been compared to measures of the
optical depth obtained from observations towards the LMC and Galactic
bulge (\cite{alc97a}; \cite{ren97}; \cite{uda94}; \cite{ala95};
\cite{alc97b}), with further tentative comparisons also starting to
emerge from observations towards SMC (\cite{alc97c}) and from pixel
experiments directed towards M31 (\cite{ans97}; \cite{cro97}). For
non-pixel based experiments, observational determinations of the
optical depth are based on the model-independent estimate
   \be
      \tau_{\rm meas} = \frac{1}{E} \frac{\pi}{4} \sum_{i=1}^{N_{\rm obs}}
      \frac{t_{{\rm e},i}}{{\cal E}(t_{{\rm e},i})}, \label{e3}
   \ee
({\em e.g.}\/ \cite{uda94}; \cite{alc97a}; \cite{alc97b}), where
$t_{{\rm e},i}$ $(i = 1 \dots N_{\rm obs})$ are the measured event
timescales (we define $t_{\rm e} \equiv 2R_{\rm e}/V_{\rm T}$, with
$V_{\rm T}$ the MACHO velocity across the observer--source line of
sight), ${\cal E}$ is the efficiency with which timescales $t_{\rm e}$
are detected and $E$ is the ``effective exposure''; that is the
average observation time per source star multiplied by the total
number of stars observed. Equation~(\ref{e3}) basically measures the
fraction of the total observing time for which microlensing events are
in progress. Errors are typically determined by a boot-strap
method in which the range in optical depth is estimated from random
timescale realisations generated from the observed $t_{{\rm e},i}$
(e.g. \cite{alc97b}; \cite{alc97a}). Han \& Gould (\cite{han95}) have
shown that error estimates based on naive Poisson statistics can
significantly underestimate the true error.

Alcock et al. (1997a) point out that equation~(\ref{e3}) is {\em
not}\/ a measure of the total optical depth, but only of the optical
depth of events which fall within a particular range of timescales
(those for which ${\cal E} > 0$). They stress that an estimate of the
total optical depth requires one to input a timescale distribution.

In this study we define the concept of {\em observable}\/ optical
depth and use it, together with the MACHO collaboration's 2-year LMC
results, to place limits on the total optical depth, and hence MACHO
halo fraction, for the ``standard'' isothermal halo model analysed by
MACHO. The analysis highlights the dependency of the results on MACHO
mass for a given Galactic model. Whilst optical depth measurements
allow one to determine a lower limit on the density of MACHOs for a
given Galactic distribution function, no upper limit can be obtained
without using extra information, such as can be obtained from a
rate-timescale analysis.

\section{Optical depth-timescale distribution} \label{s2}

Because of the timescale dependence of equation~(\ref{e3}), its 
theoretical analogue is {\em not}\/ equation~(\ref{e1}), since this is
an implicit integral over all event timescales. Instead, one can use
the following expression:
   \be
      \tau = \frac{\pi}{4} \int_{t_{\rm e}({\cal E} > 0)} t_{\rm e}
      \frac{{\rm d}\Gamma}{{\rm d}t_{\rm e}} \, {\rm d}t_{\rm e},
      \label{e4}
   \ee
where $\Gamma$ is the event rate. Equation~(\ref{e4}) is almost a
re-statement, in somewhat expanded form, of the relation $\tau =
(\pi/4) \langle t_{\rm e} \rangle \Gamma$, where $\langle t_{\rm e}
\rangle$ is the average event duration. The one difference is the
restriction to timescales $t_{\rm e}({\cal E} > 0)$.

Equation~(\ref{e4}) points to an expression for the differential
contribution to the optical depth from events of duration $t_{\rm
e}$:
   \be
      \frac{{\rm d}\tau}{{\rm d}t_{\rm e}} = \frac{\pi}{4} t_{\rm e}
      \frac{{\rm d}\Gamma}{{\rm d}t_{\rm e}}. \label{e5}
   \ee
Evidently, just as the rate-timescale distribution ${\rm d}\Gamma/{\rm
d}t_{\rm e}$ depends upon the spatial, velocity and MACHO mass
distributions, so too must the optical depth-timescale distribution
${\rm d}\tau/{\rm d}t_{\rm e}$. Its integral, if performed over a
restricted range of timescales, is similarly model dependent, unlike
the expression in equation~(\ref{e1}). Hence, to evaluate $\tau$
within a certain range of event durations $t_{\rm e}$ one must specify
both the full Galactic distribution function and the MACHO mass
function.

As an example, we employ the cored isothermal halo model originally
analysed by Griest (\cite{grie91}), and denoted model~S in the MACHO
collaboration's halo analyses. For the case of a discrete mass
function and stationary line of sight, the rate-timescale distribution
towards the LMC for this model is 
   \be
      \frac{{\rm d}\Gamma}{{\rm d}t_{\rm e}} = \frac{2 \, V_{\rm
      c}^2\rho_0}{m} (a^2 + R_0^2) \int_0^L \frac{ \beta(m,x)^2
      \exp[-\beta(m,x)]}{(x^2 -2xR_0 \cos b \cos l +a^2 + R_0^2)} \,
      {\rm d}x \label{e6}
   \ee
(c.f. Griest \cite{grie91}), where $V_{\rm c} = 220$~km~s$^{-1}$ is
the halo velocity normalisation, $\rho_0 = 0.0079~\sm$~pc$^{-3}$ is
the local halo density, $a = 5$~kpc is the halo core radius, $R_0 =
8.5$~kpc is the Sun's Galactocentric distance, ($l = 280\arcdeg$, $b =
-33\arcdeg$, $L = 50$~kpc) is the LMC position in Galactic coordinates
and $\beta \equiv (2R_{\rm e}/V_{\rm c} t_{\rm e})^2$. From
equations~(\ref{e5}) and (\ref{e6}), the optical depth-timescale
distribution becomes
   \be
      \frac{{\rm d}\tau}{{\rm d}t_{\rm e}} = \frac{\pi \, V_{\rm c}^2
      \rho_0 t_{\rm e}}{2 m} (a^2 + R_0^2) \int_0^L \frac{
      \beta(m,x)^2 \exp[-\beta(m,x)] }{(x^2 -2xR_0 \cos b \cos l +a^2
      + R_0^2)} \, {\rm d}x. \label{e7}
   \ee

\begin{figure}
\epsfig{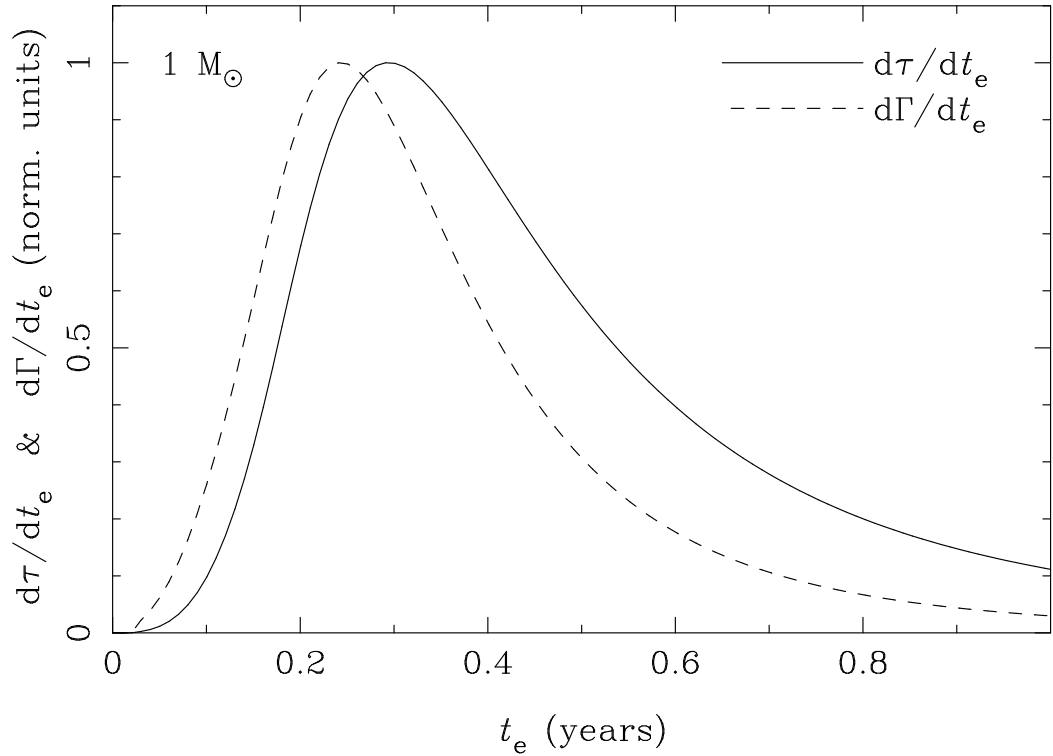}
\caption{The optical depth-timescale and rate-timescale distributions
towards the LMC for a standard isothermal halo and a lens mass $m =
1~\sm$. The line of sight is assumed to be stationary with respect to
the Galactic rest frame. The peak of both distributions is normalised
to unity for ease of comparison.}
\label{f1}
\end{figure}
Both distributions are plotted in Figure~(\ref{f1}) for the LMC
direction, assuming a MACHO mass $m = 1~\sm$. To ease comparison, the
distributions are both normalised such that their peak values are
unity. It is evident from the figure just how much more sensitive the
optical depth is than the rate to longer duration events.

\section{Observable and observed optical depths} \label{s3}

Having defined the optical depth-timescale distribution we can now
compare theoretical prediction with observation in either
of two ways. The first method would be to directly apply
equations~(\ref{e3}) and (\ref{e4}) to microlensing data. Instead, we
proceed by defining the {\em observable}\/ optical depth
   \be
      \tau_{\rm oble} \equiv \int_{t_{\rm e}({\cal E} > 0)} {\cal
      E}(t_{\rm e}) \frac{{\rm d}\tau}{{\rm d}t_{\rm e}} \, {\rm
      d}t_{\rm e}. \label{e8}
   \ee
This quantity represents the total optical depth which is potentially
observable to a microlensing experiment with detection efficiency
${\cal E}$. Since it already incorporates the detection efficiency, it
should be compared not to equation~(\ref{e3}) but to the directly {\em
observed}\/ optical depth in the absence of efficiency corrections:
   \be
      \tau_{\rm obsd} = \frac{1}{E} \frac{\pi}{4} \sum_{i=1}^{N_{\rm obs}}
      t_{{\rm e},i}. \label{e9}
   \ee
In the limit of low-number statistics, which is presently the case for
searches towards the LMC and SMC, comparison of the quantities
$\tau_{\rm oble}$ and $\tau_{\rm obsd}$ should provide a more stable
estimate than can be obtained from equations~(\ref{e3}) and (\ref{e4}),
since it is less dependent on the efficiency estimates for particular
timescales.

An evaluation of the quantities $\tau_{\rm oble}$ and $\tau_{\rm
obsd}$ allows a straightforward estimate of the halo fraction $f =
\tau_{\rm obsd}/\tau_{\rm oble}$ for our adopted halo model. This
fraction represents the total MACHO fraction, not just the fraction
within some timescale range. The first 2 years of MACHO observations
towards the LMC have uncovered 8 microlensing candidates
(\cite{alc97a}), including a likely binary event (\cite{ben96}). MACHO
analyses both this sample and a six-event sub-sample which excludes
the binary event, since it seems likely that the event originates from
within the LMC, and also excludes one of the weaker of the other
candidates to preserve the overall average event duration. We restrict
our analysis to this 6-event sub-sample. The timescales of these
6 events yield an observed optical depth $\tau_{\rm obsd} = 5.7 \times
10^{-8}$ for an effective exposure $E = 1.82\times
10^7$~star-years. Note that this is smaller than the value quoted by
MACHO since it does not compensate for the effect of
efficiencies. Instead, it represents the optical depth actually
measured by the experiment.

As mentioned previously, the potentially observable optical depth
$\tau_{\rm oble}$ is sensitive not only to the halo model but to the
assumed MACHO mass function. We calculate inferred halo fractions for
discrete MACHO mass functions with masses ranging from $0.01 -
10~\sm$. This range includes the $0.1 - 1~\sm$ range favoured by the
MACHO collaboration's maximum-likelihood analysis of the
rate-timescale distribution (\cite{alc97a}). The MACHO 2-year
detection efficiencies are incorporated as required in
equation~(\ref{e8}).

\begin{figure}
\epsfig{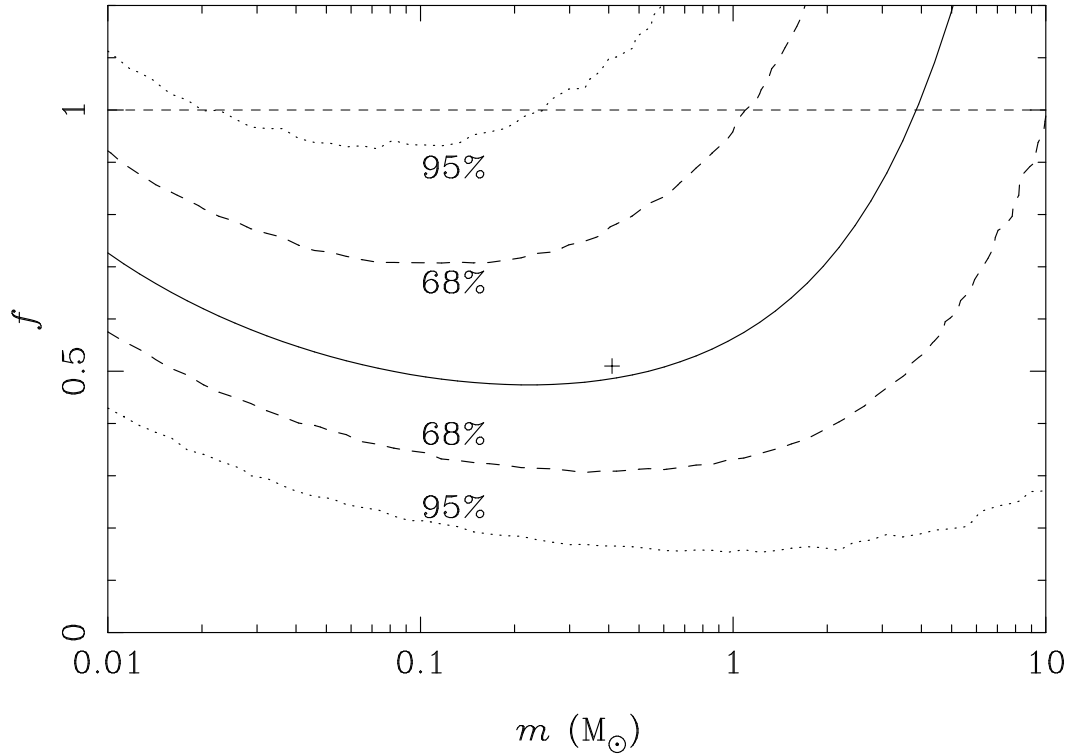}
\caption{Halo fraction determinations resulting from the 2-year MACHO
LMC 6-event sample. A discrete MACHO mass function is assumed. The
solid line denotes the preferred value as a function of $m$. Also
shown are $68\%$ and $95\%$ confidence-level regions bounded by dashed
and dotted lines, respectively. The horizontal dashed line at $f = 1$
demarcates a full MACHO halo, whilst the cross denotes the MACHO
collaboration's 2-D maximum likelihood solution based on the
rate-timescale distribution ($m = 0.41~\sm$, $f = 0.51$).}
\label{f2}
\end{figure}

The solid line in Figure~(\ref{f2}) shows the preferred halo fraction
$f$ as a function of assumed MACHO mass $m$. The errors on $f$ are
determined by Monte-Carlo simulation and are discussed in the
following section. The most obvious point to note is the mass
dependency of $f$. Whilst the preferred value for $f$ is quite stable
between $\sim 0.1 - 1~\sm$ at $f \simeq 0.5$, it becomes arbitrarily
large for small and large MACHO masses, with a minimum preferred value
of $f = 0.47$ occurring for $m = 0.23~\sm$. This variation is directly
related to the timescale correspondence between the efficiency and
optical depth distributions. For very high- or low-mass MACHOs the
peak in the optical depth-timescale distribution occurs at relatively
long and short timescales, respectively, where the efficiency is
low. Hence, $\tau_{\rm oble}$ is very small for these cases, and so $f
\propto \tau_{\rm oble}^{-1}$ becomes large for a given $\tau_{\rm
obsd}$. Hence, for a particular Galactic distribution function,
optical depth measurements alone can specify $f$ only as a function of
$m$.

\section{Error estimates} \label{s4}

To obtain error estimates on our inferred halo fraction $f$ we use a
Monte-Carlo procedure in which 10000 microlensing ``experiments'' are
conducted for each assumed MACHO mass $m$ and halo fraction $f$. The
errors are obtained from the resulting distribution of ``observed''
optical depths.

The calculation procedure is as follows: for each assumed mass $m$ the
efficiency-corrected rate-timescale distribution ${\cal E} {\rm
d}\Gamma/{\rm d}t_{\rm e}$ and optical depth-timescale distribution
${\cal E} {\rm d}\tau/{\rm d}t_{\rm e}$ are calculated for a full
MACHO halo $(f = 1)$, together with their integrals $\Gamma_{\rm
oble}$ and $\tau_{\rm oble}$. Comparison of $\tau_{\rm oble}$ with
$\tau_{\rm obsd} = 5.7 \times 10^{-8}$, as computed from the MACHO
2-year LMC results, gives the preferred estimate of $f(m)$ discussed
in the previous section. For a range of assumed $f$, the expected
number of detectable events for the model is computed as $N_{\rm exp}(m) = f
E \Gamma_{\rm oble}$. This number is used to generate a Poisson
realisation $N_{{\rm obs},j}$ for mass $m$, fraction $f$ and
experiment $j$. Event durations $t_{{\rm e},i}$ $(i = 1 \dots
N_{{\rm obs},j})$ are generated from the distribution ${\cal E} {\rm
d}\Gamma/{\rm d}t_{\rm e}$. The optical depth observed in experiment
$j$, $\tau_j$, is then computed from equation~(\ref{e9}). Repeating
this process for each experiment results in the distribution
$P_{\tau_j}(f,m)$, of which some fraction $F_{\tau_j > \tau_{\rm
obsd}}(f,m)$ give $\tau_j$ in excess of the measured value of
$5.7 \times 10^{-8}$. For a given $(f,m)$, the quantity $\langle
\tau_j \rangle$, the average over all experiments, is
found to be typically in agreement to within $0.5\%$ of the quantity
$f \tau_{\rm oble}$ as obtained by direct integration over ${\cal E}
{\rm d}\tau/{\rm d}t_{\rm e}$.

For an assumed $m$, the values of $f$ for which $F_{\tau_j > \tau_{\rm
obsd}}$ is 0.16 and 0.84 (0.025 and 0.975) bound a $68\%$ ($95\%$)
confidence region in $f$. These confidence intervals are bracketed by
the dashed (dotted) lines in Figure~(\ref{f2}). It is clear that these
intervals are larger for larger lens masses $m$. This is because a
measured $\tau_{\rm obsd}$ implies a fixed total time $\sum_{i} t_{{\rm
e},i}$ for the summed event durations.  The number of events required
to produce this summed duration is inevitably larger for low-mass
MACHOs than for high-mass MACHOs, due to the smaller Einstein radius
of low-mass MACHOs [{\em c.f.}\/ equation~(\ref{e2})]. As a result,
$N_{\rm exp}$ is required to be large for low-mass MACHOs so Poisson
fluctuations about $N_{\rm exp}$ are small. The converse is true for
very massive MACHOs, which give rise to inherently longer
durations. Since $N_{\rm exp}$ for these objects is typically smaller,
larger Poisson fluctuations can arise, which in turn produce larger
errors in $f$.

The actual number of detected events, $N_{\rm obs}$, is of course a
known quantity but is not directly used in the optical depth
analysis. This is because the analysis is concerned only with the sum
of the event timescales $\sum_{i} t_{{\rm e},i}$ not the number of
events, and this is essentially why the analysis places no constraint
on the MACHO mass $m$. Rate-timescale analyses use the observed number
of events together with the probability of observing each event
duration (for an assumed underlying timescale distribution) to
constrain both the MACHO density and mass.

Figure~(\ref{f2}) shows that, using optical depth measurements alone,
one can obtain firm mass-independent lower limits on $f$ for the
assumed halo model. The measured value for $\tau_{\rm obsd} = 5.7
\times 10^{-8}$ implies lower limits of $f > 0.3$ ($84\%$ confidence)
and $f > 0.15$ ($97.5\%$ confidence) for the model adopted here. Both
of these estimates are in excellent agreement with the MACHO
collaboration's model-independent boot-strap determination
(\cite{alc97a}), and thus support MACHO's assertion that the lower
limits it derives ought to be robust. Figure~(\ref{f2}) also
highlights the fact that upper limits cannot be placed without extra
information on the MACHO mass. The MACHO collaboration's boot-strap
analysis gives upper limits which are consistent with the minimum
derived upper-limit values in Figure~(\ref{f2}). These minimum values
are $f < 0.7$ ($84\%$ confidence) and $f < 0.9$ ($97.5\%$
confidence). Note that these upper limits apply only to specific MACHO
masses, and that the mass at which the upper limit minimises is a
function of the required confidence level. The comparison of
upper-limit estimates confirms the MACHO collaboration's concerns that
its model-independent boot-strap method may tend to underestimate
upper limits, since it does not take account of the possible
contribution of events with durations exceeding the experiment lifetime.

Plotted in Figure~(\ref{f2}) is the MACHO collaboration's preferred
value for $f$ and $m$ based on a 2-D maximum likelihood analysis of
the rate-timescale distribution (\cite{alc97a}). The correspondence
between this estimate ($f = 0.51$, $m = 0.41~\sm$) and the optical
depth $f(m)$ constraint for the same mass is reassuring and shows that
optical depth measures can also serve as important consistency checks
on rate-timescale analyses. One would not have the right to expect
such good agreement in cases where the assumed model is a poor
approximation of the actual Galactic distribution function, or where
the underlying timescale distribution is poorly sampled by the
experiment. For $m = 0.41~\sm$ we obtain $f = 0.48^{+0.62}_{-0.31}$,
where the quoted errors bound a $95\%$ confidence interval. This
implies a {\em total}\/ halo optical depth $\tau =
2.24^{+2.91}_{-1.45} \times 10^{-7}$, which is to be compared to the
MACHO collaboration's own model-independent estimate of
$\tau_{2}^{200} = 2.06^{+2.38}_{-1.29} \times 10^{-7}$ for events with
durations between 2 and 200 days (\cite{alc97a}). Comparison of the
two values shows that the MACHO collaboration's estimate is slightly
lower than ours, which is to be expected since our estimate is not
restricted to a certain timescale range. However, the MACHO estimate
is not {\em much}\/ lower because, for the assumed Galactic model, the
timescales produced by $0.41~\sm$ lenses typically fall within the
MACHO efficiency range and are thus relatively well
sampled.

\section{Discussion} \label{s5}

The optical depth is a familiar concept in gravitational microlensing
studies and is widely used as a means to determine relatively
model-independent constraints on the density of MACHOs in our
Galaxy. In this study we emphasise that a comparison between the
observed optical depth and the predicted total optical depth is a
model-dependent procedure and that failure to take proper account of
this may give rise to misleading results.

We have shown that even if one specifies a distribution function for
the MACHO population, optical depth constraints on their density will
still be a function of the MACHO mass. Whilst optical depth
measurements can be used to place firm mass-independent lower limits
on the MACHO density they cannot be used to place upper limits. The
rate-timescale distribution, however, can provide both lower and upper
limits on the MACHO density because of its ability to simultaneously
constrain both the MACHO density and typical mass.

In cases where more than one Galactic component contributes
significantly to the observed lensing rate, as is believed to be the
case for searches directed towards the Galactic bulge, a proper
calculation of the relative contribution of each component to the
observed optical depth requires one to assume both distribution
functions and mass functions for each component. Such assumptions are
implicit in calculations which simply compare observed and theoretical
optical depths without regard to timescale distributions. This is also
relevant to LMC searches if the LMC or an intervening structure is
contributing significantly to the microlensing statistics
(\cite{sahu94}; \cite{zhao97}; \cite{zar97}). Similarly, if one wishes
to make a comparison between the halo fractions of different Galactic
models, as inferred from optical depth measurements, on must specify
full distribution functions and mass functions for each model.

In summary, optical depth measurements can provide firm lower limits
on the density in MACHOs and can serve as useful consistency checks
for rate-timescale analyses. By themselves, optical depth measurements
cannot provide upper limits on the MACHO density. Comparisons between
the preferred MACHO density of different Galactic components, or
of different Galactic models, are meaningful only if one assumes
both a full distribution function and MACHO mass function for each
component or model.

\acknowledgments
The author wishes to thank Will Sutherland for providing the MACHO
2-year LMC efficiencies and David Bennett for pointing out errors in an
earlier version of this work. This research is supported by an EU
Marie Curie TMR postdoctoral fellowship.

\end{document}